\documentclass{article}
\usepackage{algorithm}
\usepackage{algpseudocode}
\usepackage{amsmath} 
\usepackage[dvipsnames]{xcolor} 
\usepackage{listings}         
\usepackage{tcolorbox}      
\tcbuselibrary{listings} 
\usepackage{graphicx}
\usepackage{booktabs}
\usepackage{multirow}   
\usepackage{caption}

\usepackage{tabularx}
\usepackage{booktabs}

\newcolumntype{C}{>{\centering\arraybackslash}X}

\usepackage[preprint]{neurips_2024}

\usepackage[utf8]{inputenc} 
\usepackage[T1]{fontenc}    
\usepackage{hyperref}       
\usepackage{url}            
\usepackage{booktabs}       
\usepackage{amsfonts}       
\usepackage{nicefrac}       
\usepackage{microtype}      
\usepackage{xcolor}         
\usepackage{natbib}
\usepackage{color}

\newcommand{\todo}[1]{}
\renewcommand{\todo}[1]{{\color{red} TODO: {#1}}}

\title{Protect$^*$: Steerable Retrosynthesis through Neuro-Symbolic State Encoding}

\author{%
  \textbf{Shreyas Vinaya Sathyanarayana\textsuperscript{1}, 
  Shah Rahil Kirankumar\textsuperscript{1},} \\
  \textbf{Sharanabasava D. Hiremath\textsuperscript{1,2,3}, 
  Bharath Ramsundar\textsuperscript{1}} \\
  \textsuperscript{1}Deep Forest Sciences \\
  \textsuperscript{2}Departament de Farmacologia, Toxicologia i Química Terapèutica, Universitat de Barcelona \\
  \textsuperscript{3}Institut de Nanociència i Nanotecnologia IN2UB, Universitat de Barcelona \\
  \texttt{\{shreyas, bharath\}@deepforestsci.com} \\
}

\begin{document}

\maketitle

\begin{abstract}
 Large Language Models (LLMs) have shown remarkable potential in scientific domains like retrosynthesis; yet, they often lack the fine-grained control necessary to navigate complex problem spaces without error. 
 A critical challenge is directing an LLM to avoid specific, chemically sensitive sites on a molecule—a task where unconstrained generation can lead to invalid or undesirable synthetic pathways. 
 In this work, we introduce Protect$^*$, a neuro-symbolic framework that grounds the generative capabilities of Large Language Models (LLMs) in rigorous chemical logic.
 Our approach combines automated rule-based reasoning - using a comprehensive database of 55+ SMARTS patterns and 40+ characterized protecting groups - with the generative intuition of neural models.
 The system operates via a hybrid architecture: an ``automatic mode'' where symbolic logic deterministically identifies and guards reactive sites, and a ``human-in-the-loop mode'' that integrates expert strategic constraints.
 Through ``active state tracking,'' we inject hard symbolic constraints into the neural inference process via a dedicated protection state linked to canonical atom maps.
 We demonstrate this neuro-symbolic approach through case studies on complex natural products, including the discovery of a novel synthetic pathway for Erythromycin B, showing that grounding neural generation in symbolic logic enables reliable, expert-level autonomy.
\end{abstract}

\section{Introduction}

The integration of Large Language Models (LLMs) into scientific discovery has unlocked new capabilities, particularly in domains requiring complex reasoning like chemical synthesis. 
Frameworks such as DeepRetro~\citep{deepretro} have advanced the state-of-the-art by combining the generative power of LLMs with structured search algorithms. 
However, a significant limitation persists: the difficulty of imposing fine-grained, expert-driven constraints on the generative process. 
Existing systems often lack mechanisms to identify which molecular sites require protection or to suggest appropriate protecting groups, leading to the generation of strategically flawed synthetic pathways.
Furthermore, most current retrosynthesis models focus primarily on bond disconnection prediction but fail to account for chemoselectivity and regioselectivity issues - specifically, whether a site will react first or if the intended molecule will form as the major product. This oversight often results in chemically plausible but practically unfeasible pathways.

To address this critical gap, we introduce Protect$^*$, a neuro-symbolic framework that bridges the gap between neural intuition and symbolic validity.
Unlike purely neural approaches that must learn chemical rules from data distributions, our system leverages a hybrid architecture where explicit constraints guide neural generation.
We employ a rigorous rule-based engine grounded in over 55 SMARTS patterns to automatically infer reactive sites, and a logic-based scoring system to suggest optimal protecting groups from a library of 40+ candidates.
Crucially, these constraints are not merely "suggestions" but are enforced through a persistent Protection State.
Through our active state tracking, these symbolic constraints are injected into the neural inference context, effectively creating a "guardrail" that steers the LLM away from invalid pathways without requiring expensive model fine-tuning.

\section{Background}
The method presented in this paper is an extension of the DeepRetro~\citep{deepretro} system, a modular, hybrid framework for retrosynthetic analysis. DeepRetro is a hybrid LLM + Monte Carlo Tree Search (MCTS) based approach to generate retrosynthesis pathways. The framework is designed to be model-agnostic, integrating various LLMs (e.g., Anthropic's Claude series \citep{anthropic2024claude3}) as its primary reasoning engine to propose novel disconnections. This approach allows the LLM to creatively explore chemical space while the MCTS algorithm systematically builds and expands the synthesis tree, grounding the generative power of the LLM within a structured search process.

Recognizing that computational metrics alone cannot capture chemical feasibility or elegance, DeepRetro employs a multi-faceted evaluation pipeline. Each LLM-proposed step is rigorously validated using stability, validity and hallucination checks before being added to the search tree. While quantitative metrics like Pathway Success Rate and Top-k accuracy are used for benchmarking, the framework's philosophy emphasizes their limitations; such metrics can penalize novel or more elegant pathways not present in the ground-truth data. Consequently, DeepRetro places a strong emphasis on qualitative Case Study Analysis, where human experts assess the novelty and practical value of generated pathways. This human-in-the-loop validation is critical for navigating complex syntheses and serves as the motivation for developing more direct methods of human guidance.

\section{Methodology}
\begin{figure}
    \centering
    \includegraphics[width=1\linewidth]{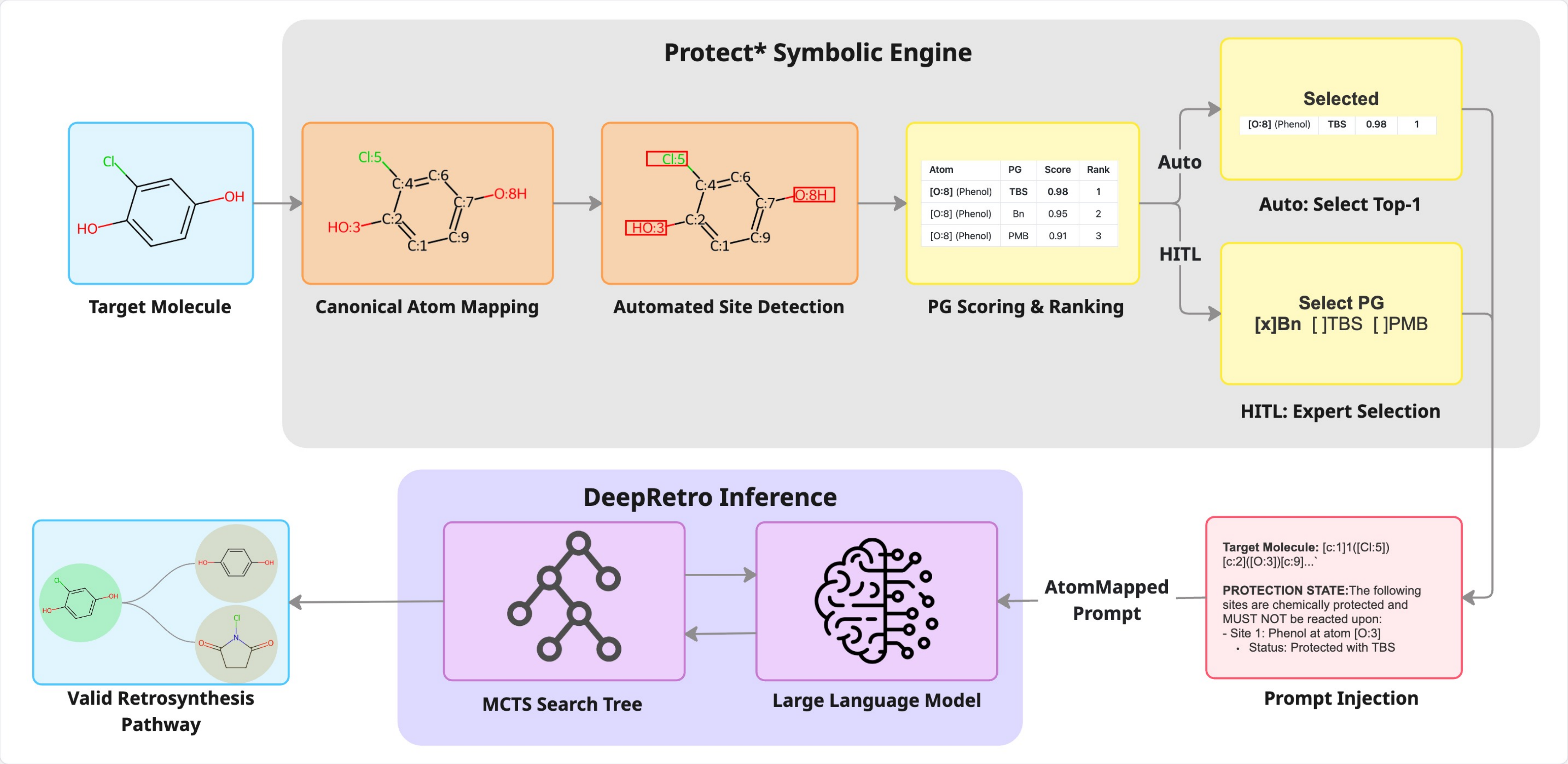}
    \caption{The Protect$^*$ architecture. The system identifies potential reaction sites in the molecule that can be protected. 
    It then suggests a list of protecting groups for each site. The chemist tags the reaction site that is to be protected. 
    This is then tracked in the persistent protection state where the framework links the protecting group to specific atom map numbers. 
    The atom-mapped SMILES along with an updated LLM prompt containing explicit constraints is then sent into the DeepRetro engine to obtain the retrosynthesis pathway.}
    \label{fig:hero_fig}
\end{figure}

A fundamental challenge in directing LLMs for retrosynthesis arises from the nature of the SMILES representation. 
As a linear string, it lacks an inherent mechanism to selectively mark specific atomic sites as non-reactive. 
This mirrors challenges in other domains; for instance, directly using LLMs as prompt encoders for diffusion models can degrade performance due to a misalignment between the model's generalist training and the task's need for discriminative features \citep{ma2024exploringrolelargelanguage}. Similarly, in our context, simply instructing a model via a prompt to ``not react at a specific functional group'' is often insufficient, as the LLM can struggle to ground such a spatial-chemical concept onto the string representation, leading to hallucinations and strategically flawed suggestions. 
To overcome this, Protect$^*$ introduces a formal mechanism that first automatically identifies protection sites via stable atom mapping, suggests appropriate protecting groups, and then enforces these constraints through prompt engineering and state tracking (Figure \ref{fig:hero_fig}).

\subsection{Automated Protection Site Detection}
The first stage of Protect$^*$ employs RDKit substructure matching on canonically atom-mapped SMILES to automatically identify functional groups that may require protection during synthesis. We maintain a comprehensive database of 55+ SMARTS patterns organized into 10 categories: alcohols (primary, secondary, tertiary, allylic, benzylic, etc.), phenols, diols and polyols, amines, heterocyclic N-H groups (indole, pyrrole, imidazole), carbonyls, carboxylic acids and derivatives, thiols, terminal alkynes, and phosphorus-containing groups. Each pattern is annotated with reactivity levels and compatible protecting groups. Using canonical atom mapping ensures that site identification is stable and deterministic across different SMILES permutations.

For each detected site, the system queries a protecting group database containing 40+ characterized groups spanning silyl ethers (TMS, TES, TBS, TIPS, TBDPS), benzyl ethers (Bn, PMB, DMB, Trt), acetals (MOM, MEM, THP), esters (Ac, Piv, Bz), carbamates (Boc, Cbz, Fmoc, Alloc), sulfonamides (Ts, Ns, SES), and specialized groups for thiols, carbonyls, and diols. 
Each protecting group entry includes protection/deprotection conditions, stability profiles (acid, base, hydrogenation), orthogonality scores, and cost metrics. 
The system ranks suggestions based on compatibility with the functional group, orthogonality with other protecting groups present in the molecule, and synthetic accessibility.

\subsection{Interaction Modes}
Protect$^*$ supports two complementary interaction modes. In \textit{automatic mode}, the system identifies all protection sites, selects the top-ranked protecting group for each site based on the scoring function, and registers them in the protection state. 
This mode is suitable for routine protection strategies and enables fully automated retrosynthesis planning. 
In \textit{human-in-the-loop (HITL) mode}, the system presents the ranked list of protecting group suggestions for each detected site, allowing the expert chemist to select based on their strategic knowledge of the overall synthesis. 
This mode is valuable for complex syntheses requiring non-obvious orthogonal protection strategies.

\subsection{Contextual Constraint Encoding}
Once protecting groups are selected (automatically or by the user), Protect$^*$ performs Contextual Constraint Encoding. The core innovation is the \texttt{ProtectionState} object - a persistent state that exists outside the neural network. This state explicitly maps specific canonical atom indices to their active protection status, serving as an immutable source of truth that survives across inference steps.

This symbolic state is interfaced with the neural model via contextual prompt injection, where we augment the inference context with a structured representation of the protection state (detailed in Appendix~\ref{sec:prompts}). 
By explicitly listing protected sites using their canonical atom map numbers (e.g., ``Atom [O:7] is protected with TBS''), we effectively bind the abstract logic of the symbolic module to the specific token representations of the neural model. 
This bridge shifts the cognitive burden from ambiguous natural language interpretation to deterministic, logic-constrained inference, ensuring that the user's strategic constraints are mathematically preserved within the generative process.

\subsection{Reducing Human Intervention}
This approach provides deterministic control over the generative process, which significantly reduces the likelihood of hallucinations at the constrained site. This reduction in model error translates directly to a decrease in required human intervention, which we quantify using partial re-runs, a procedure where an expert must halt a failing synthesis plan and restart it from an earlier, valid intermediate due to model error. As shown in Table~\ref{tab:rerun_comparison}, the complex synthesis of Erythromycin B required four such re-runs using the vanilla DeepRetro system. In contrast, by enforcing the necessary strategic constraint from the outset, the Protect\*-steered synthesis required zero re-runs. The simpler syntheses of Prostaglandin E2 and Quinine were solved without re-runs by both methods, establishing a baseline.

\begin{table}[h!]

 \centering
 \caption{Comparison of partial re-runs required for steered vs. unsteered synthesis planning. A re-run signifies a necessary human intervention due to model hallucination.}
 \label{tab:rerun_comparison}
 \begin{tabular}{@{}l c c@{}}
  \toprule
  \multirow{2}{*}{\textbf{Case Study Molecule}} & \multicolumn{2}{c}{\textbf{Partial Re-runs Required}} \\
  \cmidrule(l){2-3} 
  & \textbf{Protect$^*$} & \textbf{Vanilla DeepRetro} \\
  \midrule
  Erythromycin B   & 0 & 4 \\
  Prostaglandin E2 & 0 & 0 \\
  Quinine          & 0 & 0 \\
  \bottomrule
 \end{tabular}
 \vspace{-1em}
\end{table}

\subsection{Algorithm and Integration}
The complete Protect$^*$ workflow is formalized in Algorithm \ref{alg:steerable_prompting} and integrated as a neuro-symbolic preprocessing layer within DeepRetro. The process operates before the main LLM call to ensure that constraints - whether automatically derived or expert-specified - are embedded in the neural model's input.

\begin{algorithm}
\caption{Protect$^*$ Algorithm}
\label{alg:steerable_prompting}
\begin{algorithmic}[1]
\Statex
\Require Target molecule $M$, LLM model $L$, Mode $\in \{\text{auto}, \text{hitl}\}$, SMARTS database $\mathcal{S}$, PG database $\mathcal{G}$
\Ensure Proposed pathways $\mathcal{P}$, Explanations $\mathcal{E}$, Confidences $\mathcal{C}$
\Statex
\State \textit{$\triangleright$ Stage 1: Automated Protection Site Detection}
\State $\text{Sites} \leftarrow \text{IdentifyProtectionSites}(M, \mathcal{S})$ \Comment{RDKit SMARTS matching}
\For{each site $s \in \text{Sites}$}
    \State $\text{Suggestions}[s] \leftarrow \text{RankProtectingGroups}(s, \mathcal{G})$ \Comment{Score by compatibility, orthogonality}
\EndFor
\Statex \textit{$\triangleright$ Stage 2: PG Selection (mode-dependent)}
\If{Mode $=$ auto}
    \State $\text{Selected} \leftarrow \{s: \text{Suggestions}[s][0] \text{ for } s \in \text{Sites}\}$ \Comment{Top-ranked PG per site}
\Else
    \State $\text{Selected} \leftarrow \text{UserSelect}(\text{Suggestions})$ \Comment{Expert chooses from ranked list}
\EndIf
\Statex \textit{$\triangleright$ Stage 3: Protection State Initialization}
\State $M_{\text{mapped}} \leftarrow \text{CanonicalizeAndMap}(M)$
\State $S_{\text{active}} \leftarrow \text{InitializeProtectionState}(M_{\text{mapped}}, \text{Selected})$
\State $C_{\text{addon}} \leftarrow \text{FormatContext}(M_{\text{mapped}}, S_{\text{active}})$
\Statex \textit{$\triangleright$ Stage 4: LLM Inference}
\State prompt$_{\text{final}} \leftarrow \text{GenerateBasePrompt}(M_{\text{mapped}}) + C_{\text{addon}}$
\State response $\leftarrow L(\text{prompt}_{\text{final}})$
\Statex \textit{$\triangleright$ Parse the raw text from the LLM into a structured format}
\State $\mathcal{P}, \mathcal{E}, \mathcal{C} \leftarrow \text{ParseLLMResponse}(\text{response})$
\Statex \textit{$\triangleright$ The ProtectionState is propagated recursively to children}
\State \textbf{return} $\mathcal{P}, \mathcal{E}, \mathcal{C}$
\end{algorithmic}
\end{algorithm}
\vspace{-1em}

\section{Results}
To demonstrate the efficacy of our contextual constraint encoding method, we performed a qualitative analysis on several complex natural products known to be challenging for automated synthesis planning. The objective was not to outperform a baseline in terms of speed or success rate, but to assess whether our steering mechanism could enable the discovery of novel and chemically elegant pathways by allowing an expert to guide the LLM away from undesirable disconnections. We conducted a case study on Erythromycin B (Section \ref{sec:case_study_erythro}), Prostaglandin E2(Appendix \ref{sec:case_study_prostaglandin}), and Quinine (Appendix \ref{sec:case_study_quinine}).

Table \ref{tab:site_detection_results} summarizes the performance of our automated framework on these targets, highlighting the system's ability to correctly identify all reactive sites and recommend the appropriate protecting group (Top-1 PG Accuracy) in every instance.

\begin{table}[h!]
\centering

\begin{tabularx}{\textwidth}{@{} l C C C C @{}}
\toprule
\textbf{Molecule} & 
\textbf{Sites Detected} & 
\textbf{Total Sites} & 
\textbf{Correct Site Recommended} & 
\textbf{Top-1 PG Accuracy} \\
\midrule
Erythromycin B   & 7 & 7 & Yes & 100\% \\
Prostaglandin E2 & 4 & 4 & Yes & 100\% \\
Quinine          & 2 & 2 & Yes & 100\% \\
\bottomrule
\end{tabularx}
\caption{Protection site detection and PG suggestion evaluation.}
\label{tab:site_detection_results}
\end{table}

\subsection{Case Study: A Novel Pathway for Erythromycin B}
\label{sec:case_study_erythro}
Erythromycin B~\citep{erythromycin_total,erythromycin} is a large, polyketide macrolide antibiotic with numerous stereocenters and sensitive functional groups, making it a formidable challenge for retrosynthesis. A key difficulty is the presence of multiple hydroxyl groups with similar reactivity. Unguided generative models often propose disconnections at the most reactive sites, which may not align with a viable long-term synthetic strategy. 
In our experiment, we used our steering mechanism to protect several specific secondary alcohol moieties, which are crucial for a late-stage macrocyclization step. 

Notably, the protecting groups selected by a human expert in our earlier HITL experiments were independently recovered by the automatic mode of Protect$^*$. The system correctly identified the secondary alcohol sites requiring protection and ranked the ethoxy (OEt) protecting group as optimal, matching the expert's strategic decision. This demonstrates that the automatic solver can achieve expert-level protection strategies for complex molecules.

\begin{figure}[h!]
  \centering
  \includegraphics[width=1\textwidth]{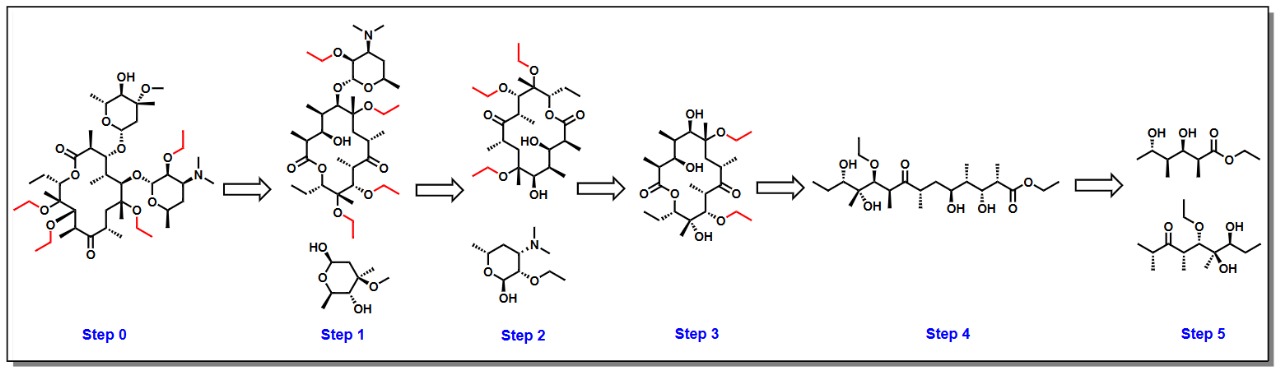}
  \caption{Retrosynthetic strategy for Erythromycin B generated by the Protect$^*$. The Hydroxyl Groups from the skeleton are protected with Ethoxy (OEt) protecting groups highlighted in red.}
  \label{fig:Erythromycin.jpeg}
\vspace{-1em}
\end{figure}

By encoding these constraints (Figure~\ref{fig:Erythromycin.jpeg}), the LLM was directed to ignore protected sites and explore alternative disconnections, yielding a chemically sound pathway.

\vspace{-1em}

\section{Conclusion}
We introduced Protect$^*$, a comprehensive framework for imposing fine-grained constraints on LLM-driven retrosynthesis. Beyond simple prompt engineering, our system provides {automated protection site detection} using 55+ SMARTS patterns and {intelligent protecting group suggestion} from a curated library of 40+ characterized groups. The dual-mode architecture - automatic for routine syntheses and human-in-the-loop for complex strategic decisions - demonstrates that automated systems can match expert-level protection strategies, as evidenced by the Erythromycin B case study.

This neuro-symbolic approach, where neural reasoning is directly constrained by structured state tracking and atom-mapped prompts, extends beyond chemistry. Variants could apply to other scientific domains with structured textual data, such as marking codons in genetic engineering or preserving architectural patterns during code generation. Unlike rigid templates, Protect$^*$ offers flexible, on-the-fly injection of both automated and expert knowledge into the generative process.

\bibliographystyle{plain}
\bibliography{ref,ref2,Retrosynthesis_zotero}

\appendix
\renewcommand{\thesection}{Appendix~\Alph{section}}
\renewcommand{\thesubsection}{\Alph{section}.\arabic{subsection}}
\section{Case Studies}
\label{sec:appendix_case_studies}

\subsection{Case Study: Strategic Bond Disconnection in Prostaglandin E2}
\label{sec:case_study_prostaglandin}
\begin{figure}[htbp]
  \centering
  \includegraphics[width=1\textwidth]{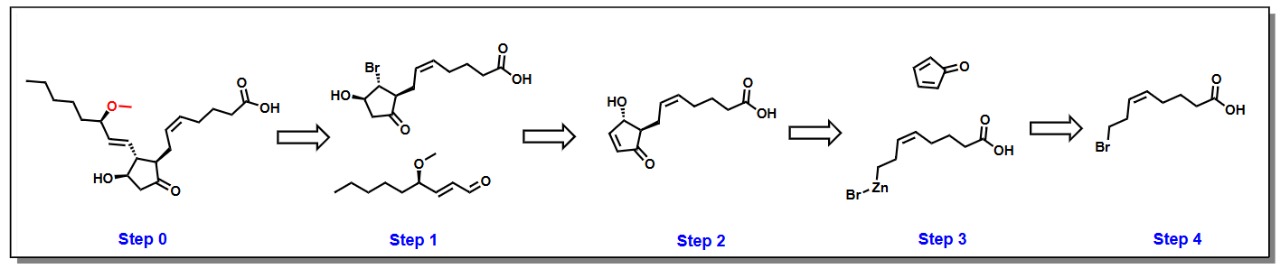}
  \caption{\textbf{Steered disconnection strategy for Prostaglandin E2.} By protecting the side chain hydroxyl, the LLM is guided away from that disconnection. The latter is protected with Methoxy (OMe) as protecting group (hightlighted in red)}
  \label{fig:prostaglandin}
\end{figure}

Prostaglandin E2~\citep{park_prostaglandin_2006} contains a delicate five-membered ring and two side chains, the disconnection of which must be carefully orchestrated. An unguided search often severs the side chains from the ring in a predictable manner. We applied our method to protect the hydroxyl on the side chain, preventing the model from pursuing disconnection on the chain. This result (see Figure~\ref{fig:prostaglandin}) demonstrates the method's utility in forcing the model to find more advanced and efficient bond-forming strategies.

\subsection{Case Study: Preserving Stereochemistry in Quinine}
\label{sec:case_study_quinine}
\begin{figure}[htbp] 
  \centering
  \includegraphics[width=1\textwidth]{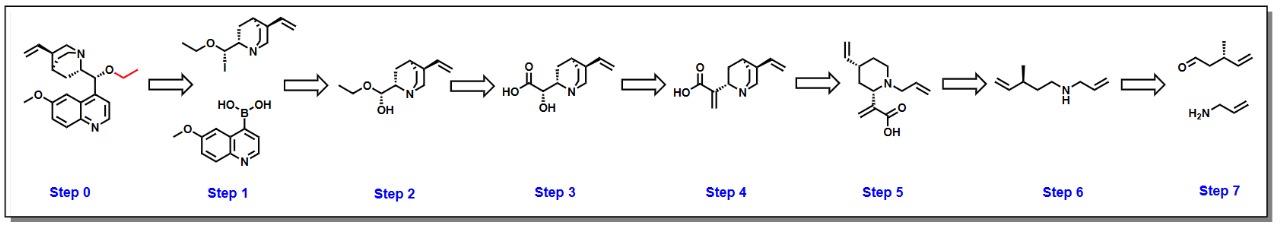}
  \caption{\textbf{Stereochemistry-preserving pathway for Quinine.} The symbolic protection of the secondary alcohol with Ethoxy (OEt) group (hightlighted in red) safeguards a critical stereocenter. The steered LLM successfully navigates the complex bicyclic structure to propose a valid disconnection at a different site, demonstrating the method's ability to enforce stereochemical constraints and discover viable pathways for stereochemically rich molecules.}
  \label{fig:quinine}
\end{figure}

The synthesis of Quinine is a classic problem in organic chemistry, complicated by its rigid bicyclic core and multiple stereocenters. A common failure mode for automated tools is proposing reactions that would compromise the delicate stereochemical integrity of the molecule. We used our contextual logical constraints to protect a key hydroxyl group, effectively signaling to the LLM that the stereocenter at that position was non-negotiable. The steered LLM successfully planned around this constraint, proposing a sequence of reactions that preserved the core stereochemistry while simplifying a different part of the molecule (see Figure~\ref{fig:quinine}). This highlights the method's potential for use in stereocontrolled synthesis, a highly complex area of chemistry.

\section{Prompts}
\label{sec:prompts}
\lstdefinestyle{promptstyle}{
  basicstyle=\ttfamily\small\color{OliveGreen},
  breaklines=true,
  postbreak=\mbox{\textcolor{red}{$\hookrightarrow$}\space},
}

\DeclareTCBListing{promptbox}{ m }{
  listing only,
  listing style=promptstyle,
  colback=Green!5!white, 
  colframe=OliveGreen,   
  boxsep=5pt,
  arc=3mm,
  title=#1,              
  fonttitle=\bfseries,
  coltitle=black,
}




\begin{figure}[H]
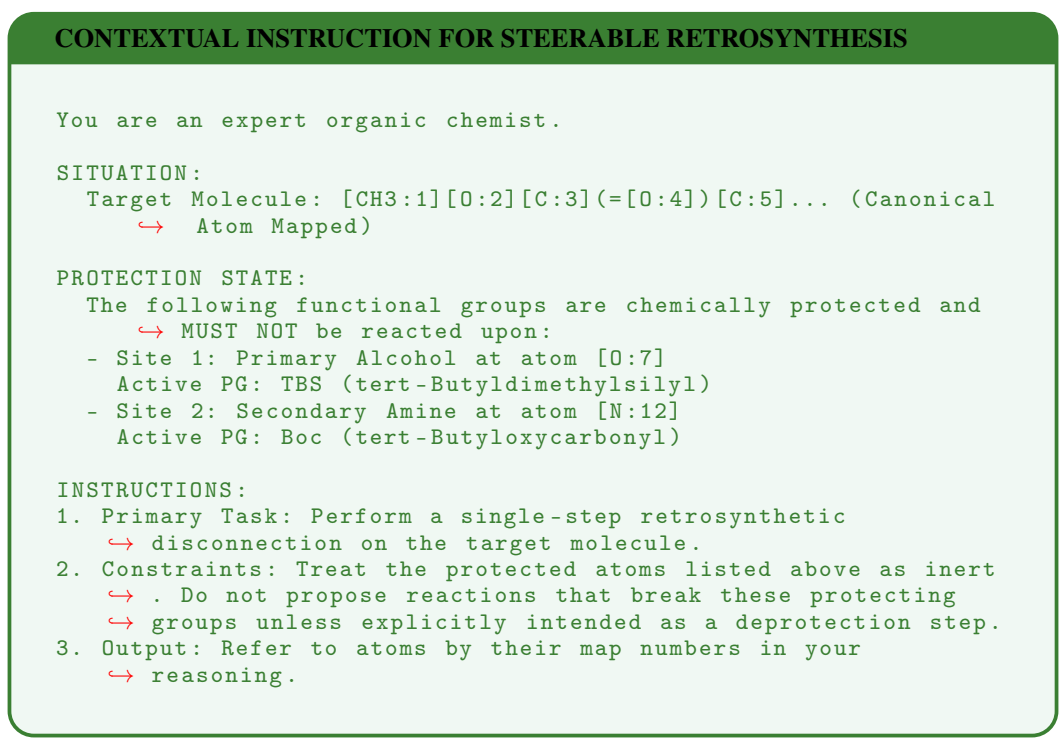

    \centering
\begin{promptbox}{CONTEXTUAL INSTRUCTION FOR STEERABLE RETROSYNTHESIS}
You are an expert organic chemist.

SITUATION:
  Target Molecule: [CH3:1][O:2][C:3](=[O:4])[C:5]... (Canonical Atom Mapped)
  
PROTECTION STATE:
  The following functional groups are chemically protected and MUST NOT be reacted upon:
  - Site 1: Primary Alcohol at atom [O:7]
    Active PG: TBS (tert-Butyldimethylsilyl)
  - Site 2: Secondary Amine at atom [N:12]
    Active PG: Boc (tert-Butyloxycarbonyl)

INSTRUCTIONS:
1. Primary Task: Perform a single-step retrosynthetic disconnection on the target molecule.
2. Constraints: Treat the protected atoms listed above as inert. Do not propose reactions that break these protecting groups unless explicitly intended as a deprotection step.
3. Output: Refer to atoms by their map numbers in your reasoning.
    \end{promptbox}
    \caption{The in-context prompt template used to guide the LLM. It provides the canonically atom-mapped SMILES and explicitly lists the protected sites by their atom map numbers and active protecting groups, instructing the model to treat them as inert.}
    \label{fig:prompt_template}
\end{figure}

\end{document}